\documentclass{article}
\usepackage{amssymb}
\usepackage{amsmath}
\usepackage{graphicx}

\title{SnIa Constraints on the event-horizon Thermodynamical model of Dark
Energy}
\author{J. Gariel\footnote{gariel@ccr.jussieu.fr}, G. Le Denmat\footnote{gele@ccr.jussieu.fr} and C. Barbachoux\footnote{barba@ccr.jussieu.fr}\\
LERMA, UMR CNRS 8112\\
Universit\'{e} P. et M. Curie ERGA, B.C. 142\\
3, Rue Galil\'{e}e, 94 200 Ivry, France}
\date{ }
\begin{document}

\maketitle

\begin{abstract}
We apply the thermodynamical model of the cosmological event
horizon of the spatially flat FLRW metrics to the study of the
recent accelerated expansion phase and to the coincidence problem.
This model, called ``ehT model" hereafter, led to a dark energy
(DE) density $\Lambda $ varying as $r^{-2},$ where $r$ is the
proper radius of the event horizon. Recently, another model
motivated by the holographic principle gave an independent
justification of the same relation between $\Lambda $ and $r$. We
probe the theoretical results of the ehT model with respect to the
SnIa observations and we compare it to the model deduced from the
holographic principle, which we call "LHG model" in the
following.Our results are in excellent agreement with the
observations for $H_{0}=64kms^{-1}Mpc^{-1}$, and $\Omega _{\Lambda
}^{0}=0.63_{-0.01}^{+0.1}$, which leads to $q_{0}=-0.445$ and
$z_{T}\simeq 0.965$.

\end{abstract}

\section{Introduction.}

Since the discovery of the presently accelereted expansion of the
universe from supernovae observations \cite{1}\cite{2}, evidences
for such an accelerated phase are increasing. The simplest
theoretical candidate to explain this acceleration is a cosmological
constant $\Lambda $. Anything producing sufficient negative pressure
- for instance a scalar field \cite{3} or a bulk viscosity \cite{4}
- could also be valid.

Before the discovery of this acceleration, phenomenological ansatze
with a variable $\Lambda (t)$ were tentatively proposed as solutions
of the cosmological ``constant" problem. For instance, laws such as
$\Lambda
(t)\thicksim t^{-2}$ \cite{5}, $\Lambda (t)\thicksim a^{-2}(t)$ \cite{6}\cite%
{7}, where $a(t)$ is the scale factor of the FLRW space-time, $\Lambda
(t)\thicksim H^{2}(t)$\cite{8} or $\Lambda (t)\thicksim \beta
H^{2}(t)+(1-\beta )H^{3}(t)H_{I}^{-1}$ \cite{9}, where $H(t)$ is the Hubble
parameter and $H_{I}$ is its exponential inflation value, were suggested.
Other proposals \cite{10}\cite{11} can also be quoted.

From a different point of view, the generalization \cite{12}
\cite{13} of the black hole and of the de Sitter event-horizon
Thermodynamics \cite{14,15} to the FLRW space-time has led to the
relation $\Lambda (t)\sim $ $r^{-2}(t)$ \cite{16} where $r$
denotes the event-horizon in the FLRW model of the universe.

Recently, this last form for $\Lambda (t)$, or, equivalently, for the dark
energy density $\rho _{\Lambda }(t)$ through $\rho _{\Lambda }(t)=\chi
^{-1}\Lambda (t)$ (with $\chi =8\pi Gc^{-4}$), has received further
confirmations based on the holographic principle \cite{17,18}.

Let us remind of the approach which can been followed to produce a model
with a time-dependent cosmological constant. We start with a type-like
perfect fluid energy-momentum tensor

\begin{equation}
T^{\alpha \beta }=\rho _{tot}u^{\alpha }u^{\beta }-P_{tot}\Delta ^{\alpha
\beta }\text{ , \ }\Delta ^{\alpha \beta }=g^{\alpha \beta }-u^{\alpha
}u^{\beta },
\end{equation}

where $u^{\alpha }$ is the 4-velocity common to all the components of the
energy density $\rho _{tot}$. We consider two components such as $\rho
_{tot}=\rho +\rho _{\Lambda }$ and $P_{tot}=P+P_{\Lambda }$. The component ($%
\rho $, $P$) is the matter with the barotropic state equation $P=(\gamma
-1)\rho $ where $\gamma $ is a constant (for instance, $\gamma =1$ for
dust). The second component is the Dark Energy (DE) with $\rho _{\Lambda }$,
the vacuum energy density, and $P_{\Lambda }$ the (negative) pressure
satisfying the state equation%
\begin{equation}
P_{\Lambda }=-\rho _{\Lambda }.
\end{equation}

Relation (2) leads to the two following alternatives:

\begin{itemize}
\item[i)] Each component is conserved separately and, of course, $\Lambda $
has to be constant.

\item[ii)] Both of the components are conserved together, $\Lambda =\Lambda
(t)$ is then possible.
\end{itemize}

The event-horizon Thermodynamics (ehT) model is derived on the
basis of point ii) by assuming an interaction between the matter
and the Dark Energy (DE hereafter). Let us remark that we write
\textquotedblleft matter" for any sort of matter except DE. Today,
the matter is the dust, the largest part of which is the Dark
Matter (DM). For sake of simplicity, we use DM to denote the dust,
encompassing the baryonic matter. In the same vein, other models
assuming an interaction between the DE and DM components of the
cosmic fluid were studied, e.g. \cite{19}.

A model such that $\Lambda \thicksim r^{-2}$ for the DE density can be used
in different ways and different contexts. For instance, in a precedent paper
\cite{16} in order to address the problem of the exit of inflation in the
early universe, we imposed as second component a perfect fluid of strings ($%
\gamma =2/3$). The model led then to $\Lambda
=3\frac{\overset{..}{a}}{a}$, which was independently considered
as an ans\"{a}tz derived by dimensional considerations by some
authors \cite{20} \cite{20bis} \cite{21}. An equivalence can be
found between the previous relation $\Lambda \thicksim
\frac{\ddot{a}}{a}$ and the forms $\Lambda \thicksim a^{-2}$ and
$\Lambda \thicksim \rho $ under specific conditions \cite{22}.

In the present paper, in order to settle some issues on the coincidence and
the recent decceleration-acceleration transition problems, we assume for the
second component a cold dark matter ($P=0$). In section 2 we review some
basic equations and relations common to the ehT and LHG models. The ehT
model is developed in section 3, particularly for the $z\leq 2$ epoch. In
section 4, in order to probe the DE assumption in this range of $z$, we
discuss how our model fits in with the type Ia supernovae observations \cite%
{23}. We deduce then the most likely values for the $H_{0}$ and $\Omega
_{\Lambda }^{0}$ parameters, as well as the decceleration parameter $q_{0}$
and the decceleration-acceleration transition redshift $z_{T}$. Finally,
sections 5 and 6 contain comments and a brief comparative discussion
concerning the results obtained by the two models.

\section{Model for $\Lambda $ and Field equations.}

In order to set the notations, we introduce some basic equations of the two
component models. The spatially flat FLRW space-time has the metric%
\begin{equation}
ds^{2}=c^{2}dt^{2}-a^{2}(t)[dR^{2}+R^{2}(d\theta ^{2}+\sin ^{2}\theta d\phi
^{2})]  \label{eq3}
\end{equation}%
where the scale factor $a(t)$ is a monotonic increasing function of the
cosmic time $t$.

We assume an universe filled by two interacting type-like perfect fluids,
namely dust (ordinary and dark matter) and Dark Energy (DE). The dust and DE
energy densities are $\rho $ and $\rho _{\Lambda }=\chi ^{-1}\Lambda $
respectively, and their corresponding pressures are $P$ and $P_{\Lambda }$.
The two state equations are $P=(\gamma -1)\rho $ with $\gamma =const,$ $0<
\gamma \leq 2$ , and $P_{\Lambda }=\omega \rho _{\Lambda }$, where $\omega $
can be variable.

We recall the field equations for the spatially flat case
\begin{equation}
3H^{2}=\chi c^{2}(\rho +\rho _{\Lambda })  \label{eq4}
\end{equation}

\begin{equation}
2\frac{\overset{..}{a}}{a}+H^{2}=-\chi c^{2}(P+P_{\Lambda )}\text{ ,}
\label{eq5}
\end{equation}
where $H\equiv \frac{\overset{.}{a}}{a}$ is the Hubble parameter,
$c$ the velocity of the light and the dot stands for the time
derivative.

Combining these two equations leads to
\begin{equation}
\dot {(H^{-1})}=\frac{3}{2}(\gamma +(1+\omega -\gamma )\Omega _{\Lambda })%
\text{ \ ,}  \label{eq6}
\end{equation}
where the dimensionless density parameter $\Omega _{\Lambda }\equiv \Lambda
c^{2}/3H^{2}$ has been introduced. The equation (6) is always valid provided
the DE is a perfect fluid.

We consider now $\Lambda $ as a vacuum energy density associated to the FLRW
event-horizon such as
\begin{equation}
\Lambda =\frac{3\alpha ^{2}}{r^{2}}\text{ \ , }  \label{eq7}
\end{equation}%
where $r$ is the proper radius of the event-horizon, and $\alpha $ is a
dimensionless constant parameter. This form of $\Lambda $ was previously
obtained by \cite{16} and \cite{17} when $\alpha =1$, and by \cite{18} when $%
\alpha \neq 1$.

Using the quantity $\Omega _{\Lambda }$, relation (\ref{eq7}) becomes
\begin{equation}
\sqrt{\Omega _{\Lambda }}=\frac{\alpha c}{rH}\text{ .}  \label{eq8}
\end{equation}%
The proper radius of the flat FLRW event-horizon is
\begin{equation}
r(t)=a(t)\int_{t}^{\infty }\frac{c\,dt^{\prime }}{a(t^{\prime })},
\label{eq9}
\end{equation}%
The derivative of (\ref{eq9}) with respect to time gives
\begin{equation}
H-\frac{\overset{\cdot }{r}}{r}=\frac{c}{r}\text{ .}  \label{eq10}
\end{equation}%
For convenience, we introduce the variable $x\equiv \ln a(t)$ such as $x=0$
today. Relation (\ref{eq10}) becomes then
\begin{equation}
1-\frac{r^{\prime }}{r}=\frac{c}{rH}=\frac{\sqrt{\Omega _{\Lambda }}}{\alpha
}\text{ \ , \ \ \ (}r^{\prime }\equiv \frac{dr}{dx}\text{),}  \label{eq11}
\end{equation}%
where the prime means the derivative with respect to $x$.

In the same manner, we can rewrite relation (\ref{eq6})
\begin{equation}
\frac{(\frac{1}{H})^{\prime }}{(\frac{1}{H})}=\frac{3}{2}(\gamma +(1+\omega
-\gamma )\Omega _{\Lambda })\text{.}  \label{eq12}
\end{equation}%
Finally, by combining equations (\ref{eq11}) and (\ref{eq12}) with the
derivative of equation (\ref{eq8}), one obtains
\begin{equation}
\Omega _{\Lambda }^{^{\prime }}=\Omega _{\Lambda }\{3[\gamma +(1+\omega
-\gamma )\Omega _{\Lambda }]-2[1-\frac{\sqrt{\Omega _{\Lambda }}}{\alpha }]\}%
\text{ \ .}  \label{eq13}
\end{equation}%
Let us emphasize that this equation is valid for any values of
$\gamma $ (constant) and $\omega $ (constant or variable),
independently of the fact that the two components $\rho $ and
$\rho _{\Lambda }$ are interacting or not.

It is useful to derive from the field equations (\ref{eq4}) and (\ref{eq5})
the decceleration parameter $q$
\begin{equation}
q\equiv \frac{-\overset{\cdot \cdot }{a}}{aH^{2}}=\frac{1}{2}[(3\gamma
-2)+3(\omega +1-\gamma )\Omega _{\Lambda }].  \label{eq14}
\end{equation}
which is valid in the two models.

In the following, we assume that the \textquotedblleft matter" component $%
\rho $ is dust ($\gamma =1$), so that (\ref{eq13}) and (\ref{eq14}) become
\begin{equation}
\Omega _{\Lambda }^{^{\prime }}=\Omega _{\Lambda }(1+2\frac{\sqrt{\Omega
_{\Lambda }}}{\alpha }+3\omega \Omega _{\Lambda })  \label{eq15}
\end{equation}%
\begin{equation}
q=\frac{1}{2}(1+3\omega \Omega _{\Lambda })\text{.}  \label{eq16}
\end{equation}%
The relations (\ref{eq3})-(\ref{eq16}) are valid in the two models
under consideration, which we denote $\Lambda (t)CDM$ models
hereafter.

From now on, the assumptions of the ehTmodel will be different
from the LHG model's ones.

\section{ Model with interacting components.}

We assume that the DE component satisfies thermodynamical state equations,
i.e. relations between its thermodynamical variables which are valid in any
space-time. Therefore, any thermodynamical state equation valid in the de
Sitter's space-time \cite{15}\cite{24} - for instance, $P_{\Lambda }=-\rho
_{\Lambda }$ and $\rho _{\Lambda }=12\pi ^{2}T_{\Lambda }^{\,2}$ ($%
T_{\Lambda }$ the temperature ) - remains valid in the FLRW
space-time. Thus, if the DE is an actual cosmological component,
its thermodynamical state equations will stay the same,
independently on the choice of the space-time as well as for any
other component. This suggests to retain the relation (\ref{eq7}) which is valid in the de Sitter's space-time when $%
\alpha =1$. In section 5, some consequences of the presence of the
parameter $\alpha $ in the ehT and LHG models are discussed. Using
the holographic principle can lead also to choose the relation
(\ref{eq7}) \cite{17}, \cite{18}. These references assume a
variable state equation ($\omega =\omega (x)$) for the DE, and
independent energy conservation laws for the matter and DE
components. Conversely, the present model assumes $\omega =-1$
(vacuum), and that the energy conservation is only valid for the
two components considered together.

Equation (\ref{eq15}) can be rewritten
\begin{equation}
\Omega _{\Lambda }^{\prime }=3\Omega _{\Lambda }(\beta _{2}-\sqrt{\Omega
_{\Lambda }})(\beta _{1}+\sqrt{\Omega _{\Lambda }})  \label{eq17}
\end{equation}%
where the constants $\beta _{1}$and $\beta _{2}$ are given by%
\begin{equation}
\beta _{1}\equiv \frac{1}{3\alpha }(\sqrt{1+3\alpha ^{2}}-1)\text{ \ , \ }%
\beta _{2}\equiv \frac{1}{3\alpha }(\sqrt{1+3\alpha ^{2}}+1)\text{, \ \ \ }%
\beta _{1},\beta _{2}>0\text{.}  \label{eq18}
\end{equation}%
By setting $\alpha =1$, Equation (17) becomes
\begin{equation}
\Omega _{\Lambda }^{\prime }=\Omega _{\Lambda }(1-\sqrt{\Omega _{\Lambda }}%
)(3\sqrt{\Omega _{\Lambda }}+1)\text{,}  \label{eq19}
\end{equation}%
which differs from Equation (\ref{eq8}) in \cite{19}. Nevertheless a
straightforward calculation (using (\ref{eq12}),(\ref{eq15}) and the
derivative of the definition of $\Omega _{\Lambda }$) gives
\begin{equation}
\Lambda ^{\prime }=2\Lambda (\sqrt{\Omega _{\Lambda }}-1)\text{ ,}
\label{eq20}
\end{equation}%
which is common to the two models. As $\Lambda ^{\prime }$ is always
negative, $\Lambda $ is decreasing with time. Observational evidences
provide a very small present value for $\rho _{\Lambda }$ (fine-tuning
problem) and of the same order as $\rho $ (coincidence problem).

Introducing the function $y(x)$ $\equiv \sqrt{\Omega _{\Lambda }}$, Relation
(\ref{eq17}) becomes%
\begin{equation}
2y^{\prime }=3y(\beta _{2}-y)(\beta _{1}+y)\text{.}  \label{eq21}
\end{equation}%
Its solution is (in the only case considered here where $y<\beta _{2}$)
\begin{equation}
K_{1}a=\frac{y^{2}}{(\beta _{2}-y)^{\frac{\alpha }{\beta _{2}\sqrt{1+3\alpha
^{2}}}}(\beta _{1}+y)^{\frac{\alpha }{\beta _{1}\sqrt{1+3\alpha ^{2}}}}}.
\label{eq22}
\end{equation}%
$K_{1}$ is a constant of integration which can be related to the initial
condition $y_{0}=\sqrt{\Omega _{\Lambda }^{0}}$. \

We derive now the expression of $r=r(y)$. Using Equations (\ref{eq11}) and (%
\ref{eq21}) yields%
\begin{equation}
d(\ln r)=dx-\frac{2dy}{3\alpha (\beta _{2}-y)(\beta _{1}+y)}.  \label{eq23}
\end{equation}%
After integration, one obtains
\begin{equation}
K_{2}r=a(\frac{\beta _{2}-y}{\beta _{1}+y})^{\frac{1}{\sqrt{1+3\alpha ^{2}}}}
\label{eq24}
\end{equation}%
or equivalently
\begin{equation}
Kr=\frac{y^{2}}{(\beta _{2}-y)^{\frac{\alpha -\beta _{2}}{\beta _{2}\sqrt{%
1+3\alpha ^{2}}}}(\beta _{1}+y)^{\frac{\alpha +\beta _{1}}{\beta _{1}\sqrt{%
1+3\alpha ^{2}}}}}\text{ \ , \ }K\equiv K_{1}K_{2}\text{.}  \label{eq25}
\end{equation}%
$K_{2}$ is a second constant of integration which depends on $y_{0}$ and $%
r_{0}=\alpha c(y_{0}H_{0})^{-1}$. The expressions of $K_{1}$ and
$K_{2}$ depend explicitly on the two priors $\Omega _{\Lambda
}^{0}$ and $H_{0}$. The current values of $\Omega _{\Lambda }^{0}$
and $H_{0}$ are $\Omega _{\Lambda }^{0}=0.7$ and
$H_{0}=72\,km.s^{-1}.Mpc^{-1}$ \cite{25}.
With these two numerical
values, it is interesting to deal with the case where $\alpha =1$
for which $\beta _{1}=\frac{1}{3}$ and $\beta _{2}=1$. One obtains
\begin{equation}
K_{1}a=\frac{y^{2}}{(1-y)^{\frac{1}{2}}(\frac{1}{3}+y)^{\frac{3}{2}}}\text{
\ , \ }K_{1}=\frac{y_{0}^{2}}{(1-y_{0})^{\frac{1}{2}}(\frac{1}{3}+y_{0})^{%
\frac{3}{2}}}=1.3686  \label{eq26}
\end{equation}%
\begin{eqnarray}
K_{2}r &=&a(\frac{1-y}{\frac{1}{3}+y}\text{ })^{\frac{1}{2}}\text{\ , or \ }%
Kr\equiv (\frac{y}{\frac{1}{3}+y})^{2}\text{, \ }r_{0}=\frac{c}{H_{0}y_{0}}%
=4980.12Mpc  \label{eq27} \\
K_{2} &=&\frac{1}{r_{0}}(\frac{1-y_{0}}{\frac{1}{3}+y_{0}}\text{ })^{\frac{1%
}{2}}=7.50265\times 10^{-5}Mpc^{-1}\text{, }K=1.02681\times 10^{-4}Mpc^{-1}.
\notag
\end{eqnarray}
However the previous values of $H_{0}$ and $\Omega _{\Lambda
}^{0}$ are model-dependent. They were obtained in the framework of
the $\Lambda CDM$ model. We shall see that starting with the same
observational SnIa data, the best fit to the $\Lambda (t)CDM$
models give appreciably different central values of $H_{0}$ and
$\Omega _{\Lambda }^{0}$.

\section{ SnIa constraints on the ehT model}

In order to compare these theoretical results with the
observations of the SnIa magnitudes, the luminosity distance
$d_{L}$ has to be expressed with respect to the redshift
$z=a^{-1}-1$. In the ehT model, it yields
\begin{equation}
d_{L}=(1+z)[(1+z)r-r_{0}]=\frac{c(1+z)}{y_{0}H_{0}}[(1+z)\frac{r}{r_{0}}-1],
\label{eq28}
\end{equation}%
where the expression of $r$ depends on $z$. As before, we only consider the
case $\alpha =1$. Both Equations (\ref{eq22}) et (\ref{eq25}) give a
parametric representation (via the \textquotedblleft parameter" $y$) of $r$
as function of $z$. Indeed, (22) yields immediately $z=z(y)$ (with $%
a=(1+z)^{-1}$).

The set of the theoretical curves ``distance moduli" $\mu $ versus the
redshift $z$,
\begin{equation}
\mu \equiv m-M=25+5\log _{10}(d_{L})\text{ , \ \ \ }\text{with }\;\;d_{L}\;\;%
\text{in Mpc},  \label{eq29}
\end{equation}
predicted by the model parametrized by the two cosmological parameters $%
y_{0}=\sqrt{\Omega _{\Lambda }^{0}}$ et $H_{0}$, can be plotted. For the two
parameters $\Omega _{\Lambda }^{0}$ and $H_{0}$ free, the best fit to the
magnitude observational data of the 157 SnIa ``Gold sample" \cite{23} can be
determined by minimizing the function $\chi ^{2}=\sum (\frac{\mu (z)-\mu
_{i}(z_{i})}{\sigma _{i}})^{2}$, where $\mu _{i}(z_{i})$ denotes the values
of the magnitude for the observational data, $\sigma_i$ the corresponding
error and the summation is taken over any of the 157 data of the sample. The
corresponding values of $\Omega _{\Lambda }^{0}$ and $H_{0}$ are derived by
numerical computation. More precisely, Equation (\ref{eq21}) is integrated
by the method of Runge-Kutta of order 4, and the expression of $z(y)$ is
deduced by use of (\ref{eq22}). With the help of Equations (\ref{eq28}) and (%
\ref{eq29}), the values of $\mu(z)$ for $z$ ranging from 0 to 100 are then
obtained. After a simple numerical evaluation of $\chi^2$ for $\Omega
_{\Lambda }^{0}$ ranging from $0$ to $1$ and $H_{0}$ from $50$ to $100$, the
best fit corresponding to $\chi ^{2}=178,7$ is obtained for
\begin{equation}
H_{0}=64^{+7}_{-4}km.s^{-1}.Mpc^{-1}, \Omega _{\Lambda
}^{0}=0.63_{-0.01}^{+0.1}\text{ ,}  \label{eq29bis}
\end{equation}%
The function $\mu (z)$ is plotted in figure 1 for $z$ ranging from
$0$ to $2$.
\begin{figure}[h]
\centerline{\includegraphics[width=2.63in,height=3.31in,angle=-90]{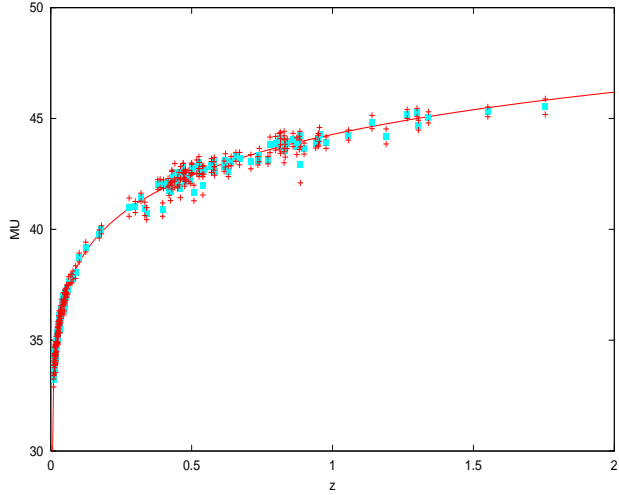}}
\caption{The ``distance moduli" $\protect\mu(z)$ of the ehT model
} \label{fig1}
\end{figure}

The likelihood function $\mathcal{L}(\Omega_{\Lambda }^{0})$ (see figure 2)
is derived by marginalization of $H_0$ and furnishes the same value of the
parameter $\Omega_\Lambda^0$.
\begin{figure}[h]
\centerline{%
\includegraphics[width=2.63in,height=3.31in,angle=-90]{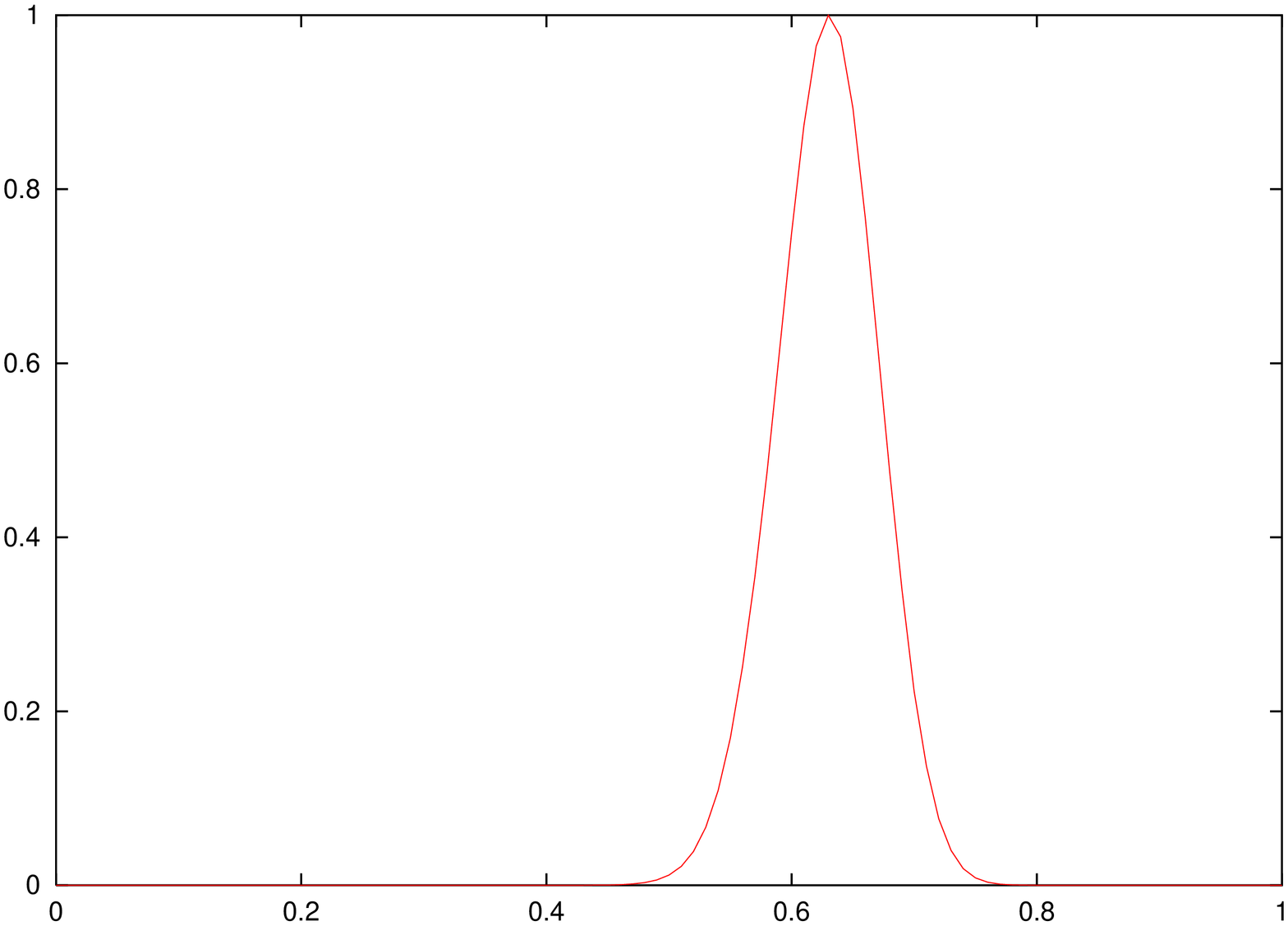}}
\caption{The likelihood function $\mathcal{L}(\Omega_{\Lambda
}^{0})$ of the ehTmodel } \label{fig2}
\end{figure}

Finally, the decceleration parameter $q$ can be expressed as a function of $%
y $ in the ehT model ( for $\alpha =1$) from equation (16) ( with $\omega
=-1 $)%
\begin{equation}
q=\frac{1}{2}(1-3y^{2})\text{.}
\end{equation}

In figure 3 the curve $q(z)$ of the ehT model is plotted. Today the
decceleration is $q_{0}=-0.445$ , and the decceleration-acceleration
transition occured at $z_{T}\simeq 0.965$.

\begin{figure}[h]
\centerline{\includegraphics[width=2.63in,height=2.31in]{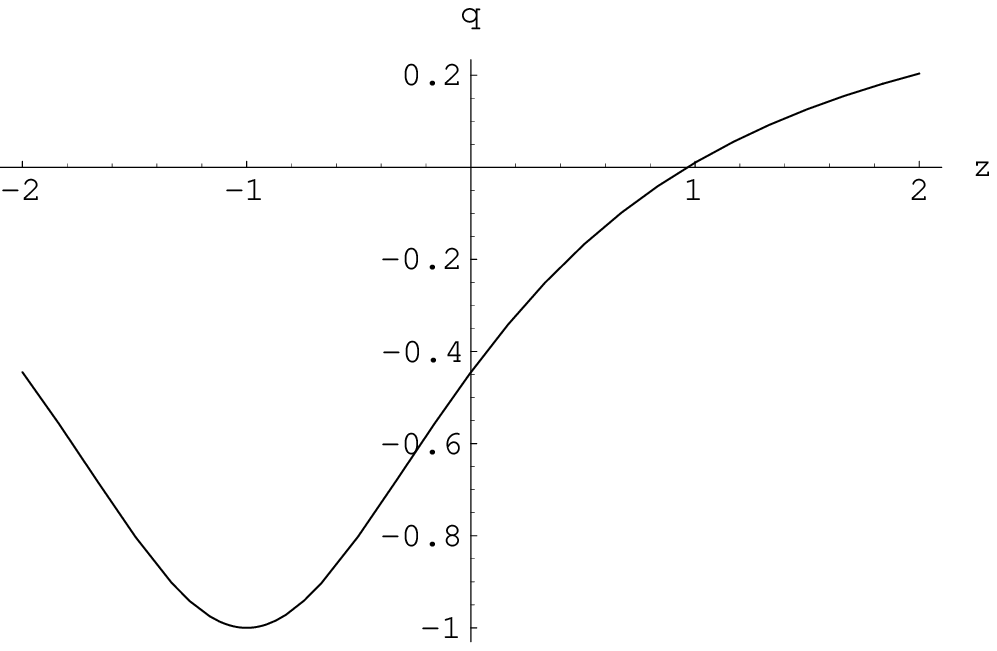}}
\caption{The decceleration parameter $q(z)$ of the ehTmodel}
\label{fig3}
\end{figure}

\section{The event horizon and the parameter $\protect\alpha $.}

We examine here the influence of the parameter $\alpha $ on the limits of
the proper radius $r$ of the event horizon (eh) in the two models. First,
let us consider the LHG model.

By comparison with the relations (\ref{eq22}) and (\ref{eq25}) of the ehT
model, the LHG model would lead to the relations ( $a$ is given by (9) of
\cite{18} and $\ r$, not explicitly given, can be deduced from their eqs.
(6) and (9)):
\begin{equation}
Y_{0}a=\frac{y^{2}(1+y)^{\frac{\alpha }{2-\alpha }}}{(1-y)^{\frac{\alpha }{%
2+\alpha }}(\alpha +2y)^{\frac{8}{4-\alpha ^{2}}}}\text{ \ , \ }\alpha \neq 2%
\text{ , }Y_{0}=\frac{y_{0}^{2}(1+y_{0})^{\frac{\alpha }{2-\alpha }}}{%
(1-y_{0})^{\frac{\alpha }{2+\alpha }}(\alpha +2y_{0})^{\frac{8}{4-\alpha ^{2}%
}}}\text{ ,}  \label{eq31}
\end{equation}%
\begin{equation}
r=\frac{\alpha }{Y_{0}^{\frac{3}{2}}H_{0}\sqrt{1-\Omega _{\Lambda }^{0}}}%
\frac{y^{2}(1+y)^{\frac{1+\alpha }{2-\alpha }}(1-y)^{\frac{1-\alpha }{%
2+\alpha }}}{(\alpha +2y)^{\frac{12}{4-\alpha ^{2}}}}\text{.}  \label{eq32}
\end{equation}%
For $\alpha =2$, the LHG model requires to start again the
calculation
from the differential equation (\ref{eq15}) which becomes:%
\begin{equation}
2y^{\prime }=y(1-y)(1+y)^{2}\text{.}  \label{eq33}
\end{equation}%
Its integration yields%
\begin{equation}
a=\frac{(1-y_{0})^{\frac{4}{3}}(1+y_{0})^{\frac{2}{3}}}{y_{0}^{2}}\frac{y^{2}%
}{(1-y)^{\frac{4}{3}}(1+y)^{\frac{2}{3}}}\exp (\frac{8}{3}(\frac{1}{1+y}-%
\frac{1}{1+y_{0}}))\text{.}  \label{eq34}
\end{equation}%
Then,
\begin{equation}
r=\frac{2c(1-y_{0})^{2}(1+y_{0})}{H_{0}\sqrt{1-\Omega _{\Lambda }^{0}}%
y_{0}^{3}}\frac{y^{2}\exp (\frac{4}{1+y}-\frac{4}{1+y_{0}})}{(1-y)^{\frac{3}{%
2}}(1+y)^{\frac{1}{2}}}\text{.}  \label{eq35}
\end{equation}%
We can see from (\ref{eq32}) or (\ref{eq35}) that $a$ tends to
infinity when $y$ tends to $1$, for any values of $\alpha $
(positive, see (8)). But the behaviour of $r$ differs because it
depends on the parameter $\alpha $, as it can be seen from (32) and
(35). \ Three cases can be distinguished for the behaviour of $r$ in
the limit $y\rightarrow 1$:
\begin{equation}
r\rightarrow 0\text{ \ if }\alpha <1  \label{eq36}
\end{equation}%
\begin{equation}
r\rightarrow \infty \text{ \ if }\alpha >1  \label{eq37}
\end{equation}%
\begin{equation}
r\rightarrow r_{i}=cst=(\frac{2}{9})^{2}\frac{c}{H_{0}\sqrt{1-\Omega
_{\Lambda }^{0}}Y_{0}^{\frac{3}{2}}}=r_{0}(\frac{2}{9}\frac{(1+2y_{0})^{2}}{%
(1+y_{0})y_{0}})^{2}\equiv \frac{c}{H_{i}}\text{ \ if }\alpha =1\text{.}
\label{eq38}
\end{equation}%
The first two cases ( i.e. $r\rightarrow 0$ and $r\rightarrow
\infty $ ) disagree with the holographic point of view, because
they would prevent any cut-off (IR and UV respectively). In
particular, the case $\alpha <1$ seems to be proscribed because it
could not prevent the singularity formation and would correspond
to the absence of black hole formation .

The third case only ($\alpha =1$) corresponds to a de Sitter asymptotic
limit. In Equation (\ref{eq38}), the index $i$ of $H$ means exponential
``inflation". Note that the limit $\frac{r_{i}}{r_{0}}$ depends only on $%
y_{0}$, and its value is : $\frac{r_{i}}{r_{0}}=1.06813$ if we take $y_{0}=%
\sqrt{0.7}$. As $r_{0}=\frac{c}{H_{0}y_{0}}=4980.12Mpc$, $r_{i}$ is equal to
$5319.42Mpc$. The expression of $r_0$ is formally the same in the two models
and depends only on the choice of the observationnal priors $H_{0}$ et $%
y_{0} $.\ However, each model leading to slightly different adjustments of
these parameters gives slightly different values of $r_0$ and $r_i$ then.

In the case of the ehT model, for any arbitrary $\alpha $, the same
phenomenon appears and the value $\alpha =2$ does not necessitate a special
study. In the limit $y\rightarrow 1$, Equations (\ref{eq26}) and (\ref{eq27}%
) give%
\begin{equation*}
a\rightarrow \infty \text{ \ and }r\rightarrow 0\text{ \ if }\alpha < 1\text{
\ (equivalently, }\beta _{2}\text{ }>1\text{)}
\end{equation*}%
\begin{equation*}
a\rightarrow \infty \text{ \ and }r\rightarrow cst=\frac{1}{K}(\frac{3}{4}%
)^{2}=5478.13Mpc\text{ \ if }\alpha =1\text{ \ (equivalently, }\beta _{2}=1%
\text{)}
\end{equation*}%
When $\alpha > 1$, $\beta _{2}$ $<1$, then $y\rightarrow \beta _{2}$ before
reaching $1$, and $a\rightarrow \infty $ , while $r\rightarrow \infty $ for
this asymptotical limit $\beta _{2}$ of $y$. From the today observational
evaluations, $\beta _{2}$ has to be $>\sqrt{0.63}=0.79$, and so $\alpha $
$< \frac{2\sqrt{0.63}}{3\times O.63-1}=1.78$. In the future, $%
\alpha $ range from $1$ to $1.78$ will become more and more narrow, tending
to $1$, as long as the equation (17) of the model, indicating a growth of $%
\Omega _{\Lambda }$, remains valid.

Thus, the case $\alpha =1$ appears to us as the most attractive. The
corresponding de Sitter's limit is $r_{i}=5478.13Mpc$. It is a little
greater than the limit of the LHG model ($5319.42Mpc$), which means a little
weaker exponential inflation.

\section{Conclusion.}

We have seen that the form $\Lambda \thicksim r^{-2}$, clearly supported by
the holographic principle, leads, in our study, to two somewhat different
models, owing to the chosen energy conservation equation. In the ehT model, $%
\alpha =1$ and the best fit ($\chi _{\nu }^{2}=1.14$) to the
SnIa's data from the ``gold" sample \cite{23} gives us $H_{0}=64$
$km.Mpc^{-1}.s^{-1}$ and $\Omega _{\Lambda }^{0}=0.63$. If $\alpha
\neq 1$ (as in the LHG model) it is worth observing that the
$\alpha <1$ values are not very attractive because they lead to
the singularity $r\rightarrow 0$ when $\Omega _{\Lambda
}\rightarrow 1$.

For the decceleration-acceleration transition epoch we find a redshift $\
z_{T}=0.96$, a value slightly higher than the ones recently published ($%
0.28\leqslant z_{T}\leqslant 0.72$)\cite{18} \cite{23} and very sensitive to
the $\Omega _{\Lambda }^{0}$ value. Comparing the values of the cosmological
parameters in various models requires to discuss not only the choice of the
parameter $\alpha $ but also the forms or relations taken for $q(z)$ (for
instance, $q(z)=q_{0}+q_{1}z$ valid when $z\ll 1$), for $\omega (z)$, or for
$d_{L}(z)$. Besides, in a given model, one has to take into account the
energy conservation laws for DM and DE.\ In most cases, the authors assume
an energy conservation law for each component separately. Here we have
considered the more \ general situation of a global conservation of the
whole energy and, necessarily, an interaction between DM and DE.\ Such an
interaction could induce higher values for the transition redshift $z_{T}$,
as noted by Amendola et al. for models with coupling \cite{26,27}.\ Future
observations in the high redshift range could allow to discriminate between
theories with coupled components and theories with distinct conservation
laws.

\end{document}